\documentclass[fleqn,usenatbib]{mnras}

\usepackage{newtxtext,newtxmath}

\usepackage[T1]{fontenc}

\DeclareRobustCommand{\VAN}[3]{#2}
\let\VANthebibliography\thebibliography
\def\thebibliography{\DeclareRobustCommand{\VAN}[3]{##3}\VANthebibliography}

\usepackage{graphicx}	
\usepackage{amsmath}	
\usepackage{color}
\usepackage{xcolor}
\usepackage{hyperref}
\usepackage[normalem]{ulem}

\title[Short title, max. 45 characters]{Effects of nucleon-nucleon short-range correlation and symmetry energy on the evolution of newly born magnetars }

\author[C. X. Liu et al.]{
C. X. Liu,$^{1}$
T. F. Feng,$^{1}$
J. M. Dong$^{2,3}$\thanks{E-mail: dongjm07@impcas.ac.cn}
\\
$^{1}$School of Physical Sciences and Technology, Hebei University, Baoding 071000, China\\
$^{2}$Institute of Modern Physics, Chinese Academy of Sciences, Lanzhou 730000, China\\
$^{3}$School of Physics, University of Chinese Academy of Sciences, Beijing 100049, China
}

\date{Accepted XXX. Received YYY; in original form ZZZ}

\pubyear{2015}

\begin{document}
\label{firstpage}
\pagerange{\pageref{firstpage}--\pageref{lastpage}}
\maketitle

\begin{abstract}
Millisecond magnetars are widely suggested as the central engines powering hydrogen-poor superluminous supernovae (SLSNe). These magnetars primarily lose huge rotational energy through gravitational wave radiation (GWR) and magnetic dipole radiation (MDR), with MDR serving as an energy source for SLSNe. We study the evolution of the magnetar spin, magnetic inclination angle, and the resulting thermal radiative luminosity of the SLSNe, where the impacts of the nucleon-nucleon short-range correlation, the mass and initial spin of the magnetar, and the density-dependent symmetry energy of the dense nuclear matter on the evolution are discussed. The relativistic mean-field theory is employed to calculate the nuclear matter properties, and we particularly concentrate on the time- and space-dependent bulk viscosity which is crucial for the magnetic inclination angle evolution. It is found that the nucleon-nucleon short-range correlation weakens the damping of bulk viscosity of dense matter and therefore inhibits the growth of magnetic inclination angle, and it reduces the MDR (GWR) peak luminosity of a canonical magnetar by several times while it raises the peak thermal radiation luminosity of SLSNe by several times. For magnetars with nonrotating mass obviously lower than the $1.4 \, \rm M_\odot $ with slow initial rotation, the magnetic inclination angle is more likely to evolve towards 0 degrees quickly, and these magnetars are not suitable as the central engine for SLSNe. Within the “family” of FSUGarnet interaction, a stiffer symmetry energy gives a lower threshold of direct Urca process and hence gives a much larger bulk viscosity coefficient, and thus it promotes the growth of the magnetic inclination angle and the GWR for canonical stars but reduces the peak brightness of SLSNe significantly. 
\end{abstract}




\section{Introduction}
Neutron stars, as compact objects in the universe with typical mass $M\sim \, \rm 1.4M_{\odot }$ and radii $R\sim 12$km, contain dense matter in their interiors. Magnetars, highly magnetized young neutron stars, are regarded to be related to numerous intriguing astronomical phenomena, including gravitational wave radiation (Andersson \citeyear{andersson2003gravitational}; Reardon et al. \citeyear{reardon2023search}), superluminous supernovae (SLSNe) (Woosley \citeyear{woosley2010bright}; Inserra et al.  \citeyear{inserra2013super}; Yu et al. \citeyear{yu2013bright}; Chen et al. \citeyear{chen2023hydrogen}), gamma-ray bursts (Zhang \& Mészáros  \citeyear{zhang2001gamma}; Metzger et al. \citeyear{metzger2011protomagnetar}), and fast radio bursts (Zhang \citeyear{zhang2013possible}; Piro \citeyear{piro2016impact}; Kashiyama \& Murase \citeyear{kashiyama2017testing}; Metzger et al. \citeyear{metzger2017millisecond}). Among these phenomena, SLSNe has attracted considerable attention since its discovery in 2007 (Ofek et al. \citeyear{ofek2007sn}; Smith et al. \citeyear{smith2007sn}; Quimby et al. \citeyear{quimby2007sn}). It is generally believed that the magnetar acts as a central engine to power the hydrogen-poor SLSNe (Inserra et al. \citeyear{inserra2013super}; Chatzopoulos et al. \citeyear{chatzopoulos2013analytical}; Yu et al.  \citeyear{yu2017statistical}), firstly proposed by Ostriker and Gunn in 1971 (Ostriker \& Gunn \citeyear{ostriker1971pulsars}) and refined by Chevalier (\citeyear{chevalier1977sn}) and Maeda et al. (\citeyear{maeda2007unique}). In this magnetar-powered model, the huge spin energy of the central magnetar is able to convert into thermal energy to heat the ejecta, which further enhances the brightness of the SLSNe. By analyzing the light curves of SLSNe, one can infer the physical properties of the magnetars involved.  
Observations indicate that a highly magnetized pulsar in whose toroidal field could be to $\sim \, 100$ times higher than the surface magnetic field, reaching $\sim 10^{16}$ G or higher, is required to explain the luminosity and timescale of SLSNe (Stella et al. \citeyear{Stella_2005}, Makishima et al.  \citeyear{PhysRevLett.112.171102}). Several mechanisms have been proposed to explain the formation of strong magnetic fields of neutron stars, such as $\alpha-\omega$ dynamo during proto-neutron star phase (Thompson \& Duncan \citeyear{1993ApJ...408..194T}). An initial spin period in the millisecond range (Gilkis \citeyear{10.1093/mnras/stx2934}) could further trigger the convective motions and the differential rotation to allow the $\alpha-\omega$ dynamo mechanism to amplify the magnetic field up to $\sim 10^{16} \, \rm G$. Generally, this process sustains $\sim 30\, \rm s$ (Duncan \& Thompson \citeyear{1992ApJ...392L...9D}). 
Furthermore, Kasen and Bildsten (\citeyear{kasen2010supernova}) as well as Woosley (\citeyear{woosley2010bright}) have highlighted that similar parameters could be responsible for the timescale and luminosity of SLSNe, making the magnetar-powered models a competitive mechanism in explaining the SLSNe.

The dissipation of spin energy of a rapidly rotating magnetar is primarily attributed to two mechanisms: gravitational wave radiation (GWR) and magnetic dipole radiation (MDR). It has shown that for newly formed millisecond magnetars, strongly magnetized relativistic winds are generated which can carry away most of the initial spin energy in a matter of minutes, leading to a faster transfer of spin energy to the ejecta (Thompson et al. \citeyear{Thompson_2004}; Bucciantini et al. \citeyear{Bucciantini_2006}). Yet, Dall’Osso \& Stella (\citeyear{Osso_2007}) have discussed the possibility that most of this initial spin energy of magnetars is released through GWR. The MDR, in particular, serves as a significant heat source for ejecta heating (Woosley \citeyear{woosley2010bright}; Kasen \& Bildsten \citeyear{kasen2010supernova}; Inserra et al. \citeyear{inserra2013super}). As ejecta expand adiabatically outward, the internal energy within SLSNe decreases. The MDR luminosity together with the GWR luminosity depends not only on the magnetic inclination angle $\chi$ between magnetic axis and rotation axis but also on the rotating angular frequency $\Omega$. Therefore, it is crucial to understand the evolution of magnetars in order to address the corresponding light curves. Understanding the evolution of magnetars is therefore paramount to addressing the observed light curves of SLSNe.

Previous studies (Yu et al. \citeyear{yu2017statistical}; Moriya \& Tauris  \citeyear{Moriya2016}; Woosley  \citeyear{woosley2010bright}; Kasen \& Bildsten  \citeyear{kasen2010supernova}) assumed a constant magnetic inclination angle of a magnetar when examining the light curves of the SLSNe. However, for a neutron star undergoing free precession, its magnetic inclination angle undergoes evolution as a result of the damping of the inner fluid matter and strong magnetic field (Dall'Osso et al. \citeyear{dall2009early}). Subsequently, Cheng et al. (\citeyear{cheng2018probing}) calculated explicitly the evolution of both the magnetic inclination angle and angular frequency for millisecond magnetars, where the evolution of these two physical quantities is coupled together.

Pulsar magnetic inclination angles are not distributed randomly at birth (Tauris \& Manchester \citeyear{tauris1998evolution}; Rookyard et al. \citeyear{rookyard2015constraints}), and Rookyard et al. discovered a bi-modal distribution of magnetic inclination angles through observations of $\gamma$-ray emitting radio pulsars (\citeyear{rookyard2015constraints}). Such a bi-modal distribution is reproduced by theoretical calculations (Dall'Osso \& Perna \citeyear{perna2017distribution}). According to Jones (\citeyear{jones1976secular}), the magnetic inclination angle is either nearly 90 degrees at birth or rapidly increases to 90 degrees. The spin energy of neutron star including the free-body precession energy is $ E_{{\rm spin}} \simeq \frac{1}{2}I\Omega^{2}(1+\epsilon_{B} \cos^2{\chi}) $, where $\epsilon_B$ is the quadrupole ellipticity of magnetic deformation. For prolate ellipsoidal magnetars ($\epsilon_B < 0$), the magnetic inclination angle $\chi$ tends to increase towards $\pi/2$ to minimize free-body precession energy through internal viscosity dissipation. 

The evolution of the magnetic inclination angle $\chi$ can be divided into two stages: the first stage in which $^3PF_2$ neutron superfluidity in the core is absent because of high temperature inside the star, and the second stage in which $^3PF_2$ neutron superfluidity occurs as the temperature drops below the critical temperature $T_c$ (Dall'Osso et al. \citeyear{dall2009early}). During the first stage, the evolution of the magnetic inclination angle is mainly determined by the interplay between bulk viscosity of the stellar matter and GWR. The stellar temperature drops due to photon emission from the surface and neutrino emission from the interior, resulting in a reduction in bulk viscosity gradually. The second stage commences when $^3PF_2$ neutron superfluidity in the core opens, in which viscous dissipation of free-body precession due to core-crust coupling mainly drives the evolution of the magnetic inclination angle (Alpar \& Sauls \citeyear{alpar1988dynamical}). Anyway, the temperature inside the star or cooling process plays a crucial role in the magnetic inclination angle evolution as the viscosity depdends strongly on the temperature.

The stellar structure is established before the calculation of cooling and not modified thereafter. The young neutron star cools primarily via neutrino emission from stellar core. This neutrino process includes the direct Urca (DUrca) process, the modified Urca (MUrca) process, nucleon–nucleon bremsstrahlung (NNB)(Yakovlev et al. \citeyear{yakovlev1999cooling}), and the possible Cooper pair breaking and formation (PBF) process (Page et al. \citeyear{page2011rapid}). These neutrino emissivity relies on the nuclear symmetry energy of dense nuclear matter (Suleimanov et al. \citeyear{Suleimanov2011}; Yakovlev et al.  \citeyear{yakovlev2001neutrino}), because the symmetry energy controls the fraction of each species in the $\beta$-equilibrium neutron star matter apart from affecting the stellar structure (Hebeler et al. \citeyear{hebeler2010constraint}; Steiner \& Gandolfi  \citeyear{steiner2012connecting}).

\begin{figure}
    \centering
    \vspace{0.0cm}
    \includegraphics[width=0.95\linewidth]{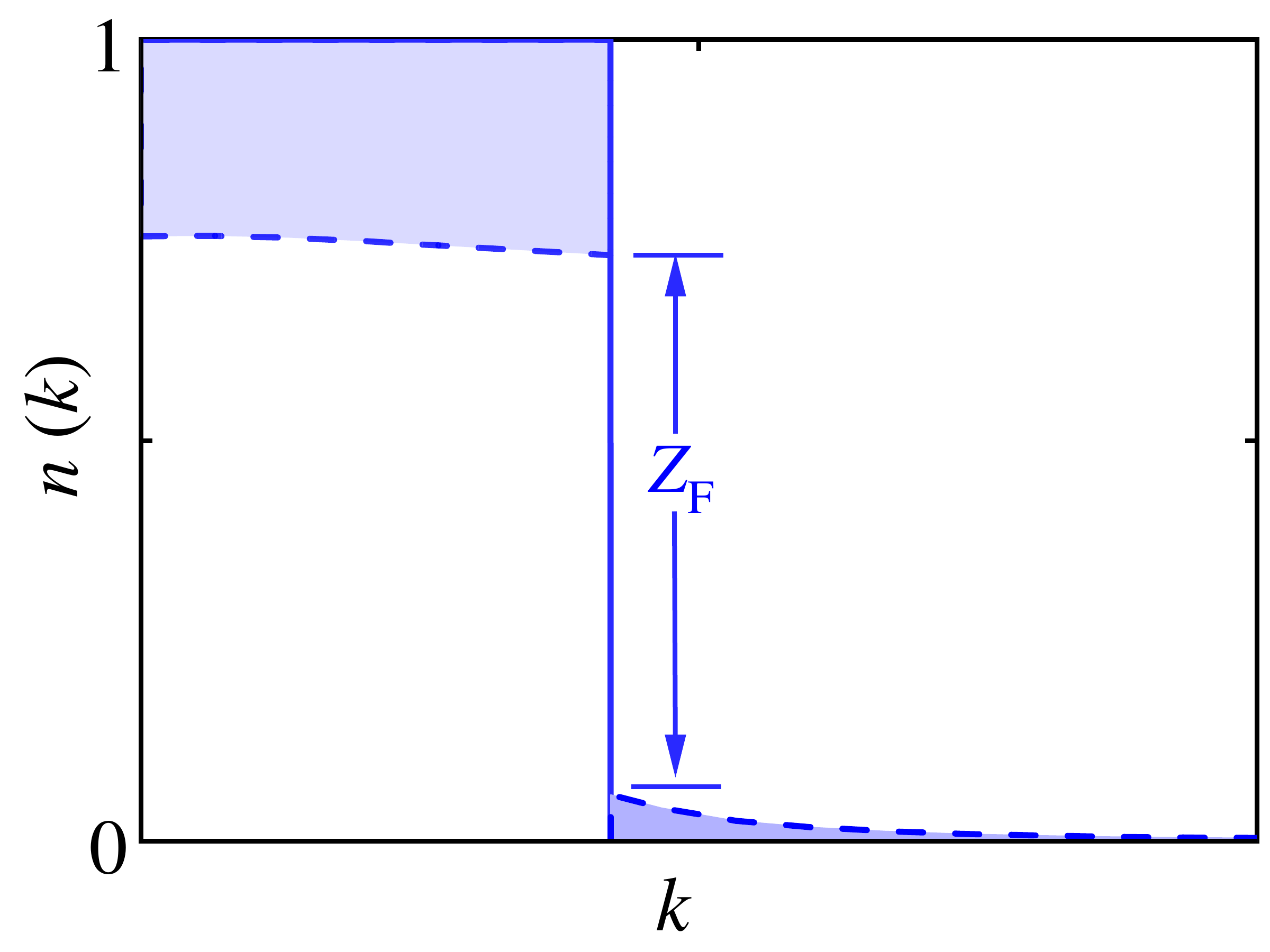}
    \caption{(Color online) Schematic representation of the effect of the Z-factor on the momentum distribution of protons or neutrons at zero temperature. }
    \label{fig: Z_factor}
\end{figure}

Both the neutrino emissivity and the bulk viscosity of neutron star matter are related to the nucleon-nucleon interaction (Haensel et al. \citeyear{haensel2000bulk}; Yakovlev et al.  \citeyear{yakovlev1999cooling}; Page et al.  \citeyear{page2004minimal}; Page et al.  \citeyear{page2006cooling}). The nucleon–nucleon correlation, in particular the short-range correlation caused by short-range repulsion and tensor interaction, is so strong that it creates a high-momentum tail, giving rise to the Fermi surface depletion (i.e. Z-factor effect). In other words, nucleons occupy energy levels above the Fermi surface, while energy levels below the Fermi surface are not fully occupied. The Z-factor at the Fermi surface is equal to the discontinuity of the occupation number in momentum distribution according to the Migdal–Luttinger theorem (Luttinger \citeyear{luttinger1960fermi}), as illustrated in Fig. \ref{fig: Z_factor}. The Z-factor reduces the level density of nucleons around the Fermi surface that controls many properties of Fermion systems related to particle-hole excitations around the Fermi energy. For instance, Dong previously investigated the effect of the Z-factor on the bulk viscosity, neutrino emissivity, and the specific heat capacity of neutron star matter (Dong \citeyear{dong2021r}; Dong et al. \citeyear{dong2016role}). Ding et al. (\citeyear{PhysRevC.94.025802}) show that the Z-factor reduces the neutron $^3PF_2$ pairing gap. Cai \& Li (\citeyear{PhysRevC.93.014619}) explored the effect of the Z-factor on the equation of state. 
In this work, we examine the impact of the Z-factor and density-dependent symmetry energy on the time- and space-dependent bulk viscosity of neutron star matter and hence on the magnetar evolution and light curves of SLSNe. 
We focus only on the first evolution stage of magnetic inclination angle evolution where the $^3PF_2$ neutron superfluidity in the core is absent. The magnetar is assumed to be made of $npe\mu$ matter, under conditions of electric neutrality and $\beta$-equilibrium. The exotic degrees of freedom, such as hyperons and quarks, are not included here. We consider the spatially non-isothermal characteristics of the stellar interior in this study during the neutron star cooling (Sales, Thiago et al. \citeyear{sales2020revisiting}). Accordingly, the temperature and the bulk viscosity are dependent on both space and time, in contrast with earlier studies that assumed a uniform temperature distribution (Woosley \citeyear{woosley2010bright}; Yu et al.  \citeyear{yu2017statistical}; Moriya \& Tauris  \citeyear{Moriya2016}; Dall'Osso et al.  \citeyear{dall2009early}; Cheng et al.  \citeyear{cheng2018probing}). 

The structure of this paper is organized as follows. We present the theoretical framework in Sec. \ref{sec: theoretical framework} about the equation of state, short-range correlation (Z-factor), bulk viscosity, and the evolution of magnetars and SLSNe. Section \ref{sec: Results and discussions} presents the results and discussions, including the effects of the Z-factor, the mass and initial spin of the magnetar, and symmetry energy on the evolution of magnetars. Finally, a summary is briefly presented in Sec. \ref{sec: Summary}.

\section{theoretical framework}
\label{sec: theoretical framework}

\subsection{Equation of state of dense matter and stellar structure}
\label{sec: eos}

The relativistic mean-field theory describes nuclear many-body systems ranging from finite nuclei to neutron stars, where the nucleon-nucleon interaction is achieved by exchanging of $\sigma, \omega, \rho$ mesons. The interacting Lagrange density is given by (Walecka \citeyear{WALECKA1974491}; Serot \& Walecka  \citeyear{Walecka1997Recent}; Fattoyev et al.  \citeyear{IUFSU2010})
\begin{flalign}\label{Lagrangian density}
   &\ \begin{aligned}
        \mathcal{L_{\rm int}} = & \Bar{\psi}[g_\sigma\sigma - (g_\omega\omega_\mu + \frac{g_\rho }{2}\boldsymbol{\tau\cdot\rho}_\mu +\frac{e}{2}(1 + \tau_3)A_\mu )\gamma^\mu ]\psi \\
        &-\frac{\kappa}{3!}(g_\sigma\sigma)^3-\frac{\lambda}{4!}(g_\sigma\sigma)^4  +\frac{\varsigma}{4!}g^4_\omega(\omega_\mu \omega^\mu )^2\\
        &+\Lambda_V(g^2_\rho\boldsymbol{\rho_\mu \rho^\mu })(g^2_\omega\omega_\mu \omega^\mu ) ,
    \end{aligned}&
\end{flalign}
where $g_\sigma$ and $g_\omega$ are the meson-nucleon coupling constants. Compared with previous relativistic mean field models, the interactions employed in this work is FSUGarnet (Chen \& Piekarewicz \citeyear{chen2015searching}), where two additional parameters $\varsigma $ and $\Lambda_{v}$ have been introduced: $\omega$ meson self-interactions as described by $\varsigma $ which softens the equation of state at high density, and the nonlinear mixed isoscalar-isovector coupling described by $\Lambda_{v}$ that modifies the density-dependence of the symmetry energy. 
The FSUGarnet provides a good description of finite nuclei, nuclear matter, and neutron star structure (Chen \& Piekarewicz \citeyear{chen2015searching}; Fattoyev et al. \citeyear{fattoyev2020gw190814}). In addition, the FSUGarnet interaction enhances the maximum neutron star mass from $1.94\ \rm M_{\odot}$ by IUFSU interaction (Fattoyev et al. \citeyear{IUFSU2010}) to $2.07 \pm 0.02 \  \rm M_{\odot}$ (Chen \& Piekarewicz \citeyear{chen2015searching}) which is in better agreement with astronomical observations of the maximum mass ($2.08^{+0.07}_{-0.07} \, \rm M_\odot$) of neutron stars (Fonseca et al. \citeyear{Fonseca_2021}).

The energy density and pressure of uniform neutron star matter under conditions of charge neutrality  and $\beta$-equilibrium are given by
\begin{flalign}\label{energy density}
     \nonumber
    &\ \begin{aligned}
        \varepsilon = &\frac{1}{\pi^2}\sum_{i=n,p}\int^{k^i_F}_0\sqrt{k^2 + (m_i -g_\sigma \sigma)^{2}}k^2dk +\frac{1}{2}m^2_\sigma\sigma^2+ \frac{\kappa}{3!}g^3_\sigma\sigma^3\\
        & + \frac{\lambda}{4!}g^4_\sigma\sigma^4 + \frac{1}{2}m^2_\omega\omega^2+\frac{\varsigma}{4!}g^4_\omega\omega^4+\frac{1}{2}m^2_\rho\rho^2+ 3\Lambda_Vg^2_\rho g^2_\omega\rho ^2\omega^2\\
        & + \frac{1}{\pi^2}\sum_{i=e,\mu}\int^{k^i_F}_0\sqrt{k^2 + m^{2}_i}k^2dk,
    \end{aligned}& \\
    &\ \begin{aligned}
        P = &\frac{1}{3\pi^2}\sum_{i=n,p}\int^{k^i_F}_0 \frac{k^4}{\sqrt{k^2+ (m_i- g_\sigma \sigma)^{2}}}dk -\frac{1}{2}m^2_\sigma\sigma^2- \frac{\kappa}{3!}g^3_\sigma\sigma^3\\ 
        & - \frac{\lambda}{4!}g^4_\sigma\sigma^4+ \frac{1}{2}m^2_\omega\omega^2+\frac{\varsigma}{4!}g^4_\omega\omega^4+\frac{1}{2}m^2_\rho\rho^2+ \Lambda_Vg^2_\rho g^2_\omega\rho ^2\omega^2\\
        & + \frac{1}{3\pi^2}\sum_{i=e,\mu}\int^{k^i_F}_0\frac{k^4}{\sqrt{k^2 + m^{2}_i}}dk.
    \end{aligned}&
\end{flalign}
The pressure provided by strong nuclear force instead of neutron degeneracy pressure plays a major role in supporting large-mass neutron stars. The widely used equation of state based on Baym-Pethick-Sutherland (BPS) (Baym et al. \citeyear{baym1971ground}) is applied for the crust.  
The mass-versus-radius relation and other relevant quantities of a static and spherically symmetric neutron stars can be derived by solving the Tolman-Oppenheimer-Volkoff equation
\begin{flalign}\label{TOV: pressure}
    \nonumber
    &\ \begin{aligned}
        \frac{dP(r)}{dr} = -\frac{Gm(r)\varepsilon(r)}{r^2}\frac{[1+ P(r)/\varepsilon(r)][1 + 4\pi r^3 P(r)/m(r)]}{1-2Gm(r)/r},
    \end{aligned} & \\
    &\ \begin{aligned}
         \frac{dm(r)}{dr} = 4\pi r^2 \varepsilon(r).
    \end{aligned} &
\end{flalign}
where $P(r)$ is the pressure of the star at distance $r$, $m(r)$ the mass inside a sphere of radius $r$, and $G$ is the gravitational constant. Given a central energy density, we integrate outward from the center until the pressure is zero, and obtain the stellar radius $R$ and the gravitational mass $m(R) = 4\pi \int_0^R r^2 \varepsilon(r) dr$.

\subsection{The nucleon-nucleon short-range correlation (Z-factor effect) and its effects on bulk viscosity}
\label{sec: Z-factor}

The strong nucleon-nucleon short-range correlation (strictly speaking, nucleon-nucleon correlation) leads to a deviation of nucleonic momentum distribution of nuclear matter from the perfect Fermi gas distribution (Dong et al. \citeyear{dong20133}), quantified by the Z-factor (0 < Z < 1). For perfect Fermi gas, its momentum distribution is described by the well-known Fermi–Dirac distribution (Z = 1). Dong et al. computed the Z-factor of $\beta$-stable neutron star matter by using Brueckner theory in combination with AV18 two-body force plus a microscopic three-body force, and the results are given by
\begin{equation}\label{Z_factor}
    \begin{aligned}
     &  Z_{F,n}(\rho) = 0.907 - 0.233\rho  - 0.480\rho^2 + 0.481\rho^3,\\
    & Z_{F,p}(\rho) = \left\{ 
		\begin{aligned}
			&0.351 + 2.332\rho,\  \rho \leq 0.15\ \rm{fm}^{-3} \\
			&0.656 + 0.451\rho \\
                    &  \; \; - 1.151\rho^2 + 0.576\rho^3,\  \rho > 0.15\ \rm{fm}^{-3}
		\end{aligned}
	\right. .
    \end{aligned}   
\end{equation}
The Z-factor is found to reduce the neutrino emissivity, the nucleons specific heat capacity, and the bulk viscosity (Dong et al. \citeyear{dong2016role}; Dong \citeyear{dong2021r}). Because the Fermi surface depletion hinders particle-hole excitation around the Fermi level. 
The Fermi-Dirac function $f(x) = 1/(1 + e^x)$ in the phase space integral being contained in the neutrino emissivity, the nucleons specific heat capacity, and the bulk viscosity requires amendments. Due to the strong degeneracy of nucleons and electrons, the primary contribution to this integral arises from the extremely narrow regions of the momentum space near the corresponding Fermi surfaces $k_F$ (Yakovlev et al. \citeyear{yakovlev2001neutrino}). Therefore, with the inclusion of the Z-factor, the momentum distribution function of nucleons near the Fermi surface at finite temperature $T$ (not quite high) is given by $Z_Ff(x)$ (Dong et al. \citeyear{dong2016role}). 

The bulk viscosity of neutron star matter is caused by several nonequilibrium neutrino emission processes, such as the DUrca process ($n \rightarrow p + l + \Bar{v}_l$, $p + l \rightarrow n + v_l$) and the MUrca processes ($n + N \rightarrow p + l + N + \Bar{v}_l $, $p + l + N \rightarrow n + N + v_l$). Here $N$ denotes neutron ($n$) or proton ($p$), and $l$ denotes electron ($e$) or muon ($\mu$). Compared to the MUrca process, the DUrca process is the most efficient, whereas it only occurs when the Fermi momentum of the proton is sufficiently high. The MUrca processes are regarded to be responsible mainly for the bulk viscosity of dense matter inside a canonical neutron star with the mass of $1.4 \, \rm M_\odot$.

The temperature is a key input for the calculations of bulk viscosity as the bulk viscosity relies on the temperature significantly (Haensel et al. \citeyear{haensel2000bulk}). Considering the fact that the stellar interior is not isothermal in the early stage (Sales, Thiago et al. \citeyear{sales2020revisiting}), the temperature $T(r,t)$ inside the star is determined by the neutron star cooling model  (\citeyear{page2006cooling}): 
\begin{flalign}\label{the cooling}
    &\ \begin{aligned}
        &\frac{d(Le^{2\phi})}{dr} = - \frac{4\pi r^2 e^\phi}{\sqrt{1 - 2Gm/c^2r}}(C_V\frac{dT}{dt} + e^\phi Q_v), \\
        &\frac{d(Te^\phi )}{dr} = - \frac{1}{\eta}\frac{Le^\phi }{4\pi r^2 \sqrt{1-2Gm/c^2 r}},
    \end{aligned}&
\end{flalign}
in the present study where $\eta$ and $\phi$ is the thermal conductivity and the red-shift, the luminosity $L$ and the temperature $T$ are both time- and space-dependent. $Q_v$ is the neutrino emissivity of all the different neutrino emission processes, which has been thoroughly studied (see, e.g., the review paper, Yakovlev et al. \citeyear{yakovlev2001neutrino}, \citeyear{yakovlev1999cooling}). After that, the effect of the Z-factor on neutrino emissivity was studied in detail by Dong et al (see the review paper, \citeyear{dong2016role}). 
The specific heat capacity $C_{V}$ is the sum of the contributions of leptons and nucleons (Page et al. \citeyear{page2004minimal})
\begin{flalign}\label{the specific heat capacity}
    &\ \begin{aligned}
        & C_{V} = \sum_{i = n, p} Z_{F,i}\frac{m_i^* p_{Fi}}{3\hbar^3} k_B^2 T + \sum_{j = \mu, e} \frac{m_j^* p_{Fj}}{3\hbar^3} k_B^2 T,\\  
    \end{aligned}&    
\end{flalign}
which takes into account the Z-factor correction, $m_i^*$ and $p_{Fi}$ are particle effective mass and the Fermi momentum. $k_B$ is the Boltzmann constant.

Currently, the initial center temperature of equation (\ref{the cooling}) is difficult to determine and relies heavily on the evolution of the protoneutron star. The protoneutron star is quite hot ($T \simeq (3\sim 6)\times 10^{11}\, \rm K$) at the beginning of its formation (Pons et al. \citeyear{Pons_2001}; Stuart L. Shapiro \citeyear{stellar_John}), and the neutrinos are trapped. Yet, nearly all the neutrinos escaped when the neutrino mean free path became comparable to the stellar radius, resulting in a significant drainage of thermal energy. The protoneutron star converts to an ordinary neutron star which is neutrino transparent. Previous studies have shown that this process lasts
about a minute, and its internal temperature drops to a few of MeV ($\sim 10^{10} \, \rm K$) (Suwa \citeyear{10.1093/pasj/pst030}, Fischer et al. \citeyear{PhysRevD.94.085012}, Cacciapaglia et al. \citeyear{PhysRevD.67.053001}, Pons et al. \citeyear{Pons_1999}, Nakazato et al. \citeyear{PhysRevC.97.035804}, Burrows \& Lattimer \citeyear{1986ApJ...307..178B}). Considering that the $\alpha-\omega $ dynamo mechanism lasts $\sim 30\, \rm s$ and note that the magnetic inclination angle starts to evolve visibly only after the strong magnetic field forms. This study focuses on the evolution of ordinary neutron stars where convection motion has stopped and are transparent to neutrinos, hence set the initial conditions of this cooling model to be $T(r=0, t =0) = 10^{10}\, \rm K$ as D. Page used. 

\begin{figure}
    \centering
    \includegraphics[width=1\linewidth]{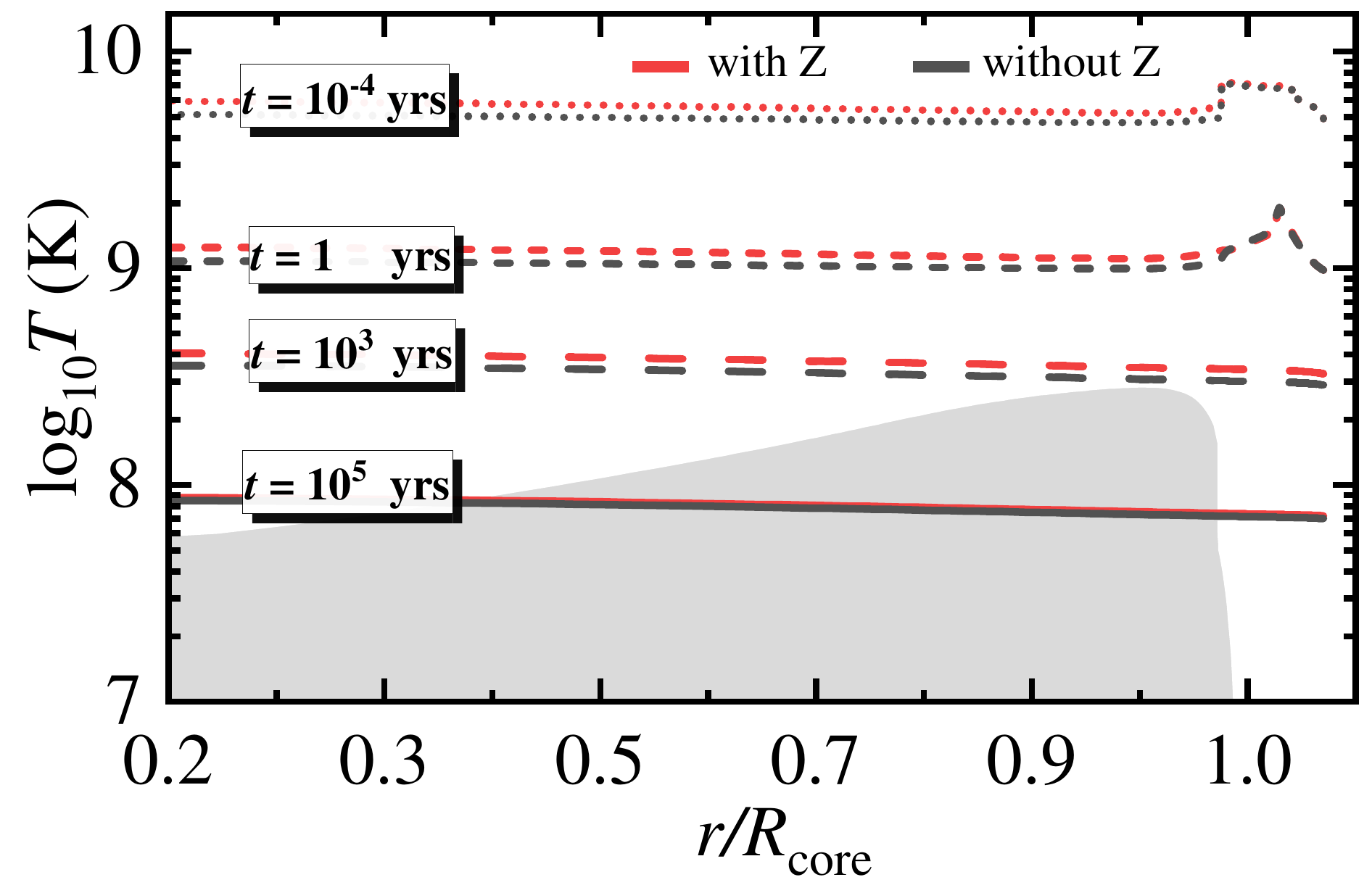}
    \caption{(Color online) The temperature $T$ inside a canonical neutron star with a non-rotating mass of $M_{{\rm TOV}} = 1.4 \, \rm M_\odot$ as a function of distance $r$. The red (black) lines represent the results with (without) the Z-factor. The gray area is the region where the $^3PF_2$ neutron superfluid with Z-factor is present, and the energy gap is given by $\Delta_{^3PF_2}(\rho) = (0.943\rho-0.050)\exp{\left[-\left(\rho/0.177\right)^{1.665}\right]}$ (Dong \citeyear{dong2021r}), as a function of nucleon number density $\rho$. The corresponding critical temperature is $k_BT_c = 0.57 \Delta$ (Page et al. \citeyear{page2004minimal}).}
    \label{fig:T_R_T}
\end{figure}

Briefly, we focus on the first stage of magnetic inclination angle evolution which is from neutrinos transparent to the critical temperature of the $^3PF_2$ neutron superfluidity. The convective movements and $^3PF_2$ neutron superfluidity in the core do not exist. The matter inside the star with a solid crust is in $\beta$-equilibrium. The cooling curve of ordinary neutron stars is calculated by using the publicly available code NScool by D. Page\footnote{\href{https://www.astroscu.unam.mx/neutrones/NSCool/}{https://www.astroscu.unam.mx/neutrones/NSCool/}}.
The calculated cooling curves are displayed in Fig. \ref{fig:T_R_T}. As mentioned above, both the nucleon-specific heat capacity and the neutrino emissivity are suppressed by the Z-factor. And it eventually slows down the cooling of the star. Moreover, in the case of canonical neutron stars ($M_{{\rm TOV}} = 1.4 \, \rm M_\odot$), $^3PF_2$ neutron superfluidity appears in the core typically in $\sim 10^3$ years, and the first stage ends.

The explicit expression for the bulk viscosity in the non-superfluid dense matter without the Z-factor effect is given as (Haensel et al. \citeyear{haensel2000bulk})
\begin{flalign} \label{ bulk viscosity of DU}
    &\ \begin{aligned}
        \xi_{l,0}^D(r,t) = &8.553 \times 10^{24} \  \frac{m_n^*}{m_n}\frac{m_p^*}{m_p} \left(\frac{\rho_e}{0.153}\right)^{1/3}T_9^4(r,t)\frac{1}{\omega_4^2}\\ 
         &\times\left( \frac{C_l}{100\ {\rm MeV}}\right)^2\Theta_{npl}\  {\rm g \, cm^{-1}\, s^{-1}},
       \end{aligned}&
\end{flalign}
\begin{flalign} \label{bulk viscosity of MU ne}
    &\ \begin{aligned}
         \xi_{n,e,0}^M(r,t) = &1.49 \times 10^{19} \left( \frac{m_n^*}{m_n}\right)^3 \frac{m_p^*}{m_p} \left( \frac{\rho_p}{0.153}\right)^{1/3}\alpha_n\beta_n\\
         &\times \left(\frac{C_e}{100\ {\rm MeV}}\right)^2 T_9^6(r,t) \omega_4^{-2} \ {\rm g \, cm^{-1} \, s^{-1}},
    \end{aligned}&
\end{flalign}
\begin{flalign} \label{ bulk viscosity of MU pe}
    &\ \begin{aligned}
        \xi_{p,e,0}^M(r,t) = &\xi_{n,e,0}^M(r,t) \left( \frac{m_p^*}{m_n^*}\right)^2 \\
        &\times\frac{(3p_{Fp} + p_{Fe} - p_{Fn})^2}{8 p_{Fp} p_{Fe}}\Theta_{pe},
    \end{aligned}&
\end{flalign}
\begin{flalign}\label{bulk viscosity of MU nu}
    &\ \begin{aligned}
        \xi_{n,\mu,0}^M(r,t) = \xi_{n,e,0}^M(r,t)\frac{p_{F\mu}}{p_{Fe}}\left(\frac{C_\mu}{C_e}\right)^2,
    \end{aligned} &
\end{flalign}
\begin{flalign}\label{bulk viscosity of MU pu} 
    &\ \begin{aligned}
        \xi_{p,\mu,0}^M(r,t) = &\xi_{n,e,0}^M(r,t) \left(\frac{C_\mu m_p^*}{C_e m_n^*}\right)^2 \\
        &\times\frac{(3 p_{Fp} + p_{F\mu} -p_{Fn})^2}{8 p_{Fp} p_{F\mu}} \frac{p_{F\mu}}{p_{Fe}} \Theta_{p\mu},
    \end{aligned}&
\end{flalign}
with $T_9 = T/(10^9 \, {\rm K})$, $\omega_4 = \omega/(10^4\, {\rm s^{-1}})$ and $C_l = 4(1-2X_p)\rho d(J(\rho))/d\rho - c^2p^2_{Fl}/3\mu_l$. $\omega = \epsilon_B \Omega \cos{\chi}$ is precession frequency. $X_p$ and $J(\rho)$ are the proton fraction and density-dependent symmetry energy, respectively. $\rho$ and $p_F$ are the particle number density and the Fermi momentum, respectively. We use $\alpha_n = 1.76 - 0.63(1.68 {\rm fm^{-1}}/k_{Fn})^2$, $\beta_n = 0.68$ from Page et al. (\citeyear{page2004minimal}). The step function $\Theta_{npl}$ is 1 if the DUrca process opens for $k_{Fn} < k_{Fl} + k_{Fp}$ ($\Theta_{npl} = 0$ otherwise), and $\Theta_{pl}$ is 1 for $k_{Fn} < (k_{Fl} + 3k_{Fp})$. 

Identically, as mentioned above, with the inclusion of the Z-factor, the bulk viscosity caused by each neutrino process is given by 
\begin{flalign}\label{the bulk viscosity with Z_factor}
    &\ \begin{aligned}
       & \xi_l^D(r,t)  = Z_{F,n} Z_{F,p} \xi_{l,0}^D(r,t),\\
       & \xi_{n,l}^M(r,t) = Z_{F,n}^3 Z_{F,p} \xi_{n,l,0}^M(r,t),\\
       & \xi_{p,l}^M(r,t)  = Z_{F,n} Z_{F,p}^3 \xi_{p,l,0}^M(r,t).
    \end{aligned}&
\end{flalign}
Owing to $0<Z<1$, the Z-factor decreases the bulk viscosity of neutron star matter. 

Finally, the total space- and time-dependent bulk viscosity of the $\beta$-stable matter in the neutron star core is
\begin{flalign}\label{the total bulk viscosity}
    &\ \xi(r,t) = \xi_{l}^D(r,t) + \xi_{n,l}^M(r,t) + \xi_{p,l}^M(r,t).&
\end{flalign}
\begin{figure}
    \centering
    \includegraphics[width=1\linewidth]{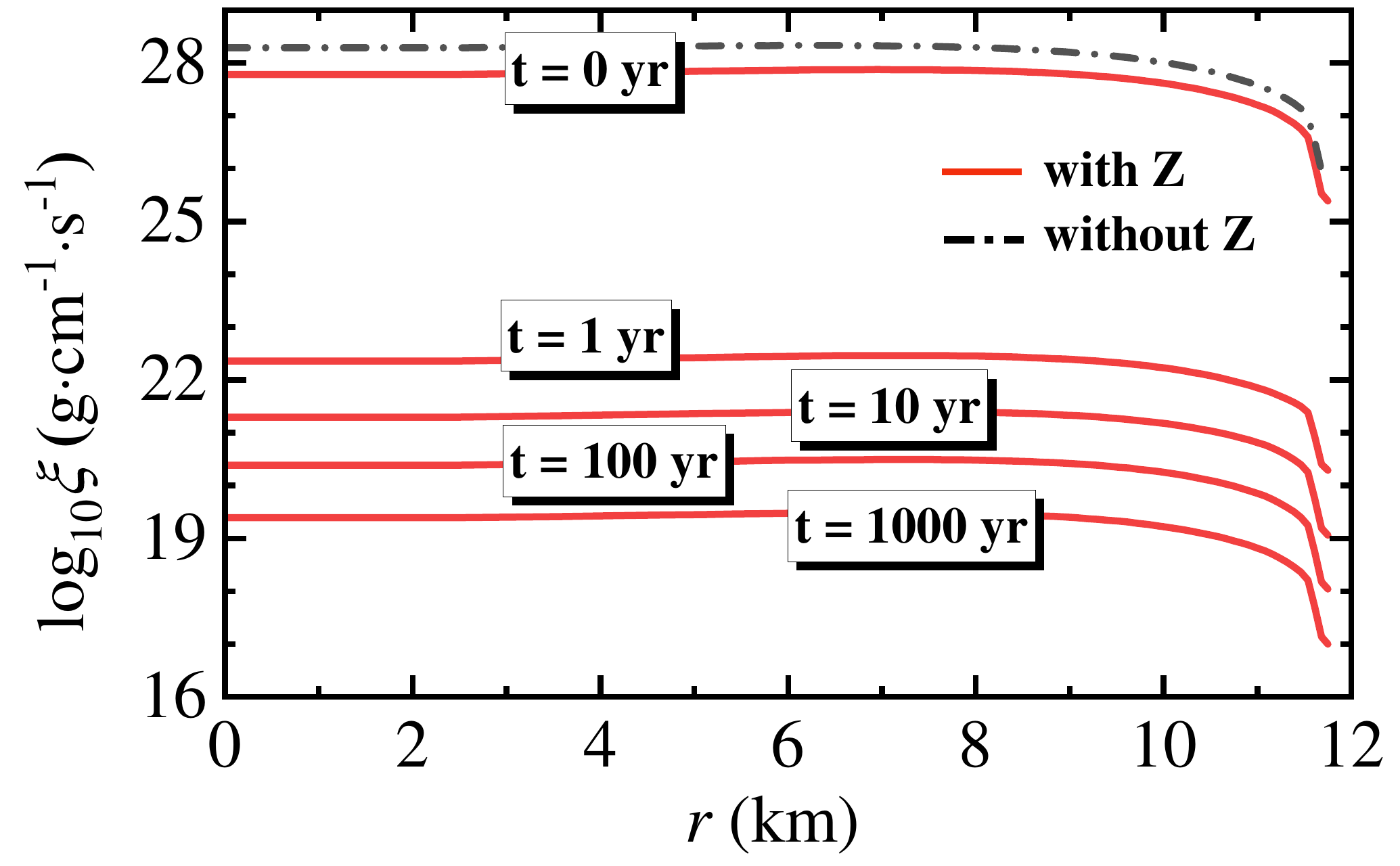}
    \caption{(Color online) The bulk viscosity $\xi$ of neutron star matter inside a canonical neutron star with a non-rotating mass of $M_{{\rm TOV}} = 1.4 \, \rm M_\odot$ as a function of distance $r$. We choose the angular frequency $\Omega = 2\pi \times 10^{3} {\rm s^{-1}}$ and an initial magnetic inclination angle $\chi  = 1 \, \rm degree$ as an illustration. The solid (dotted) lines represent the results with (without) the Z-factor. The result without the Z-factor at $t=0$ is shown for comparison.}
    \label{fig:bulk viscosity_R_T}
\end{figure}
The calculated bulk viscosity distribution inside a canonical neutron star and its evolution over time, is shown in Fig. \ref{fig:bulk viscosity_R_T} as an illustration. The figure clearly shows that at $t =0$, the Z-factor (solid line), visibly lowers the bulk viscosity (dotted black line). The bulk viscosity decreases remarkably with increasing age, as the temperature inside the star decreases during cooling process. For example, the bulk viscosity coefficient is reduced by about eight orders of magnitude at $t = 1000$ years compared with that at $t = 0$.

\subsection{Evolution of newly born magnetars and magnetar-powered models}
\label{sec: evolution}
The protoneutron star, with a temperature of tens of $10^{12}$K, is initially opaque to neutrinos and can have a radius of several tens of kilometers. Subsequently, when the neutron star becomes transparent to neutrinos, it shrinks to a radius of approximately 10 km (Metzger et al. \citeyear{metzger2011protomagnetar};  Obergaulinger \& Aloy \citeyear{obergaulinger2017protomagnetar}). This contraction leads to an increase in the spin rate due to the conservation of angular momentum.  Consequently, newly formed neutron stars may exhibit an initial spin period of at least 1 ms at birth (Metzger et al. \citeyear{metzger2011protomagnetar}; Strobel et al.  \citeyear{strobel1999properties}). Simulations of the collapse of massive progenitor have suggested that the initial spins of newly formed NSs could be around 1 ms (Heger et al. \citeyear{heger2000presupernova}; Paschalidis \& Stergioulas  \citeyear{paschalidis2017rotating}). The present work also assumes that the initial period of the neutron star is 1$\sim$3 ms.

A rapidly rotating magnetar releases its enormous spin energy primarily via MDR and GWR, and the evolution of angular frequency is written as (Cheng et al. \citeyear{cheng2015stochastic})
\begin{flalign}\label{the evolution of angular frequency}
   &\ \Dot{\Omega} = - \frac{B^2_{d} R^6\Omega^3}{6Ic^3}\sin^2{\chi} -\frac{2G\epsilon^2_BI\Omega^5}{5c^5}\sin^2{\chi}(15\sin^2{\chi} +1 )&
\end{flalign}
where $B_d$ is the surface dipole magnetic field strength at the magnetic pole. $\epsilon_B = -5\Bar{B}^2_tR^4/(6GM^2)$ represents the quadrupole ellipticity of magnetic deformation (Cutler \citeyear{cutler2002gravitational}; Dall'Osso et al.  \citeyear{dall2009early}), where $\Bar{B}_t$ is the average of the toroidal component of the magnetic field. Moriya \& Tauris (\citeyear{Moriya2016}) found that $B_d$ are of the order of $10^{14}$ G and $B_d/B_t > 0.01$ in magnetars powering SLSNe. Furthermore, this ratio is only small enough to secure a stable magnetic configuration. Consequently, in the following, we adopt typical values of $B_d = 6 \times 10^{14} $ G and $\Bar{B}_t = 5.8 \times 10^{16} $ G for newly born magnetars to show significantly the effect of Z-factor, and the decay of the magnetic field is neglected in the first stage (thousands of years at most). We calculate the moment of inertia $I$ of a neutron star by empirical fitting (Raithel et al. \citeyear{Raithel2016Model}).

According to Dall’Osso et al. (\citeyear{dall2009early}), during the first stage, the magnetic inclination angle $\chi$ of a magnetar with a liquid core evolves under the combined effect of GWR and bulk viscosity, which is expressed as
\begin{flalign}\label{the evolution of magnetic inclination angle}
   &\ \Dot{\chi} = \frac{\rm{cos}\chi}{\tau_{{\rm BV}} \rm{sin}\chi} - \frac{2G}{5c^5}I\epsilon^2_B\Omega^4\rm{sin}\chi\rm{cos}\chi(15\rm{sin}^2\chi + 1).&
\end{flalign}
The first term on the right hand side is the influence of the bulk viscosity of stellar matter on the damping of the free-body precession with damping timescale $\tau_{{\rm BV}} = 2E_{\rm {pre}}/\Dot{E}_{\rm {diss}}$, drives the $\chi$ to 90 degrees. The free-body precession energy is given by (Dall’Osso et al. \citeyear{dall2009early}) $ \ E_{{\rm pre}} \simeq  - \frac{1}{2}I\Omega^2\epsilon_B \cos^2{\chi}$ in the case of $ \epsilon_B < 0$. $\Dot{E}_{{\rm diss}} = \int \xi(r,t) \left| \nabla \cdot \boldsymbol{\delta v} \right|^2 dV$ is the bulk viscosity damping rate, where $\boldsymbol{\delta v}$ is the velocity perturbation and $\xi(r,t)$ is the time- and space-dependent bulk viscosity coefficient. The second term denotes the quench of $\chi$ driven by GWR. The radiation torque causes the spin and magnetic axes to align. The first stage of the magnetic inclination angle evolution ends when the stellar temperature reaches the $^3PF_2$ neutron superfluid critical temperature $T_c$.  

The millisecond magnetar, as a candidate of the central engine of the hydrogen-poor SLSNe, primarily dissipates its huge spin energy through the MDR and GWR where the corresponding luminosity are given by 
\begin{flalign}\label{the luminosity of GW and MDR}
    &\ \begin{aligned}
        & L_{{\rm MDR}} = \frac{B_d^2R^6\Omega^4}{6c^3}\sin^2{\chi},\\
        & L_{{\rm GWR}} = \frac{2G\epsilon_B^2 I^2 \Omega^6}{5c^5}\sin^2{\chi}(15\sin^2{\chi}+ 1).
    \end{aligned}&
\end{flalign}
The energy released by the MDR can be used to heat the ejecta of the SLSNe. In the magnetar-powered scenario, the internal energy $E_{\rm int}$ evolution equation for SLSNe can be written as (Kasen \& Bildsten \citeyear{kasen2010supernova}; Yu et al.  \citeyear{yu2015rapidly}; Ho \citeyear{ho2016gravitational}):
\begin{flalign}
    &\ \begin{aligned}
        \frac{dE_{{\rm int}}}{dt} = - \frac{E_{{\rm int}}}{3V} \frac{dV}{dt} - \frac{E_{{\rm int}}c}{\tau_{\rm o} R_{{\rm ej}}} + L_{{\rm MDR}},
    \end{aligned}&
\end{flalign}
where $V$ is the volume of the SLSNe ejecta. The second term on the right-hand side of the above equation is the thermal radiation luminosity $L_{\text{th}}$ of SLSNe. Early in evolution, the ejecta was optically thick; in other words, the optical depth $\tau_{\rm o} = 3\kappa M_{\rm ej}/(4\pi R^2_{\rm ej}) \gg 1$ with $\kappa$, $M_{\rm ej}$, $R_{\rm ej}$ representing the opacity, ejecta mass, ejecta radius, respectively. As the ejecta adiabatically expands with increasing radius, the ejecta will become optically thin ($\tau_{\rm o} \simeq 1$). The dynamical evolution of the SLSNe ejecta can generally be determined by the following equations (Kasen \& Bildsten \citeyear{kasen2010supernova}; Yu et al.  \citeyear{yu2015rapidly}),
\begin{flalign}
    &\ \begin{aligned}
        &\frac{dR_{{\rm ej}}}{dt} = v_{{\rm ej}},\\
        & \frac{dv_{{\rm ej}}}{dt} =  \frac{4\pi R^2_{{\rm ej}}E_{{\rm int}}}{3VM_{{\rm ej}}},
    \end{aligned}&
\end{flalign}
where $v_{{\rm ej}}$ is the expansion velocity of the SLSNe ejecta. We solve equations (\ref{the cooling}, \ref{the evolution of angular frequency}, \ref{the evolution of magnetic inclination angle}) to obtain the magnetar evolution in the first stage firstly and then explore the evolution of light curve of SLSNe.

\section{Results and discussions}
\label{sec: Results and discussions}
In the following calculations, we adopt the typical values $M_{{\rm ej}} = 5 \ {\rm M_{\odot}}$, $\kappa = 0.2 \ {\rm cm^2\,g^{-1}}$, $R_{{\rm ej,0}} = 3\times 10^8 \ {\rm cm}$, $v_{{\rm ej,0}} = 10^9 \ {\rm cm\,s^{-1}}$, and $E_{{\rm int,0}} = 10^{51} \ {\rm erg}$ for the ejecta mass, opacity, initial radius, initial velocity, and initial internal energy of the SLSNe, respectively (Kasen \& Bildsten \citeyear{kasen2010supernova}; Ho \citeyear{ho2016gravitational}). For the center temperature of a neutron star, this study is set to $T(r=0, t =0) = 10^{10}\, \rm K$.

\subsection{Effects of the short-range correlation (Z-factor effect)}
\label{sec: Effect of Z-factor}
\begin{figure*}
    \centering
    \includegraphics[width=1\linewidth]{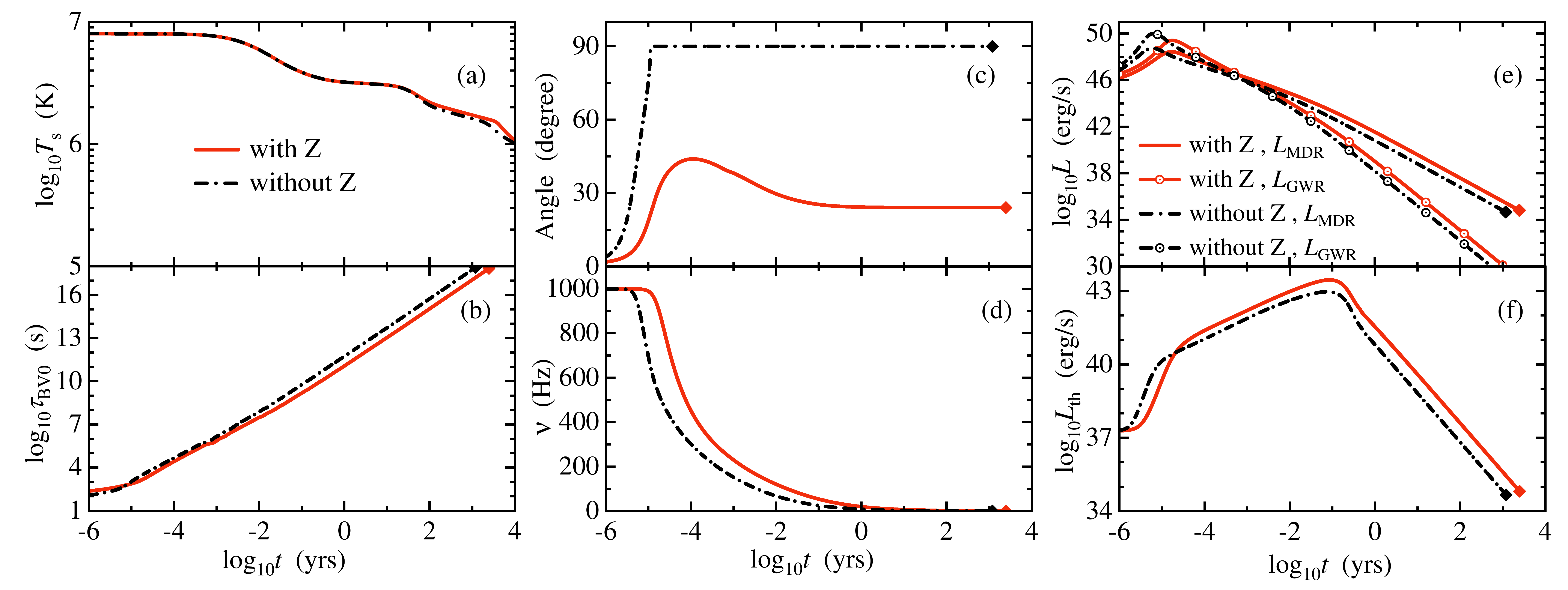}
    \caption{(Color online) The Z-factor effect on the evolution of various physical quantities of a canonical magnetar with $M_{{\rm TOV}} = 1.4 \, \rm M_\odot$ and an initial period $P_0 = 1 \, \rm ms$. The diamond symbol stands for the age that the stellar temperature reaches superfluid critical temperature $T = T_c$ (i.e., the end of the first stage). The red straight line (black dotted line) represents the calculated results with (without) the inclusion of the Z-factor. (a): Evolution of the stellar surface temperature $T_{\rm s}$; (b): Evolution of the reduced damping timescale $\tau_{\rm BV0} = \tau_{\rm BV} \sin^2{\chi } (1 + 3 \cos^2{\chi})/\cos^2{\chi}$ without the term of the magnetic inclination angle; (c): Evolution of the magnetic inclination angle $\chi$; (d): Evolution of spin frequency $\nu$; (e): Evolution of the GWR luminosity $L_{\rm GW}$ and MDR luminosity $L_{\rm MDR}$ of the magnetar; (f): Evolution of the radiated thermal luminosity $L_{\rm th}$.}
    \label{fig: T_bulk viscosity_Z}
\end{figure*}

 The equation of state obtained from the relativistic mean-field theory with FSUGarnet interaction is employed to calculate the properties of neutron star matter and to build the stellar structure. The central engine of SLSNe is a canonical magnetar with a non-rotating mass of $M_{{\rm TOV}} = 1.4\ {\rm M_{\odot}}$ and an initial period of $P_0 = 1\ {\rm ms}$, as in (Cheng et al. \citeyear{cheng2018probing}). The role of the Z-factor in the evolution of the magnetar and in the light curves of SLSNe are shown in Fig. \ref{fig: T_bulk viscosity_Z}. 

The Fermi surface depletion induced by the nucleon-nucleon correlation, known as the Z-factor, results in a reduction of neutrino emissivity and nucleonic-specific heat. Consequently, the Z-factor slows down the cooling of young magnetars. Due to the thermal relaxation process inside the magnetar, the surface temperature remains unaffected until approximately $\sim 100$ years later (Sales, Thiago et al. \citeyear{sales2020revisiting}; Lattimer et al. \citeyear{lattimer1994rapid}). As depicted in Fig. \ref{fig: T_bulk viscosity_Z} (a), the inclusion of the Z-factor leads to a slight elevation in the neutron star surface temperature after around 100 years of birth. Yet, the internal temperature of the star with the inclusion of the Z-factor is higher than that without the Z-factor as soon as neutron star formation. As a result, the Z-factor affects the bulk viscosity in three ways. Firstly, the Z-factor directly reduces the bulk viscosity coefficient and thus enlarges the viscosity timescale $\tau_{\text{BV}}$ as described in equation (\ref{the bulk viscosity with Z_factor}). Secondly, it slows down the cooling of young neutron stars and thus indirectly decreases the $\tau_{\text{BV}}$ according to equations (\ref{ bulk viscosity of DU}-\ref{bulk viscosity of MU pu}). 
Thirdly, it also affects the free-body precession of the neutron star, which in turn affects the $\tau_{\text{BV}}$, where $\tau_{\text{BV}} \sim \cos^2{\chi }/(\sin^2{\chi}\Omega^2(1 + 3 \cos^2{\chi}))$. To show the corresponding competition distinctly, $\tau_{\text{BV0}}$ of Fig. \ref{fig: T_bulk viscosity_Z} (b) does not include the magnetic inclination angle.  
It is observed that the Z-factor increases the reduced damping timescale $\tau_{\text{BV0}}$ over the first $7.5 \times 10^{-6}$ years, indicating that the first (direct) factor is dominant because the cooling process during such a short time is not distinct. The calculated $\tau_{\text{BV0}}$ (or $\tau_{\text{BV}}$) with the Z-factor is 2.5 times larger than that without the Z-factor at the initial stage. 
After $7.5 \times 10^{-6}$ years, the second (indirect) factor and the slowdown of spin-down become dominant of reducing $\tau_{\rm BV0}$ by the Z-factor and grow even more substantial as the magnetar ages until the first stage ends.

When the magnetar cooling and the corresponding damping timescale evolution are incorporated into equations (\ref{the evolution of angular frequency}, \ref{the evolution of magnetic inclination angle}), the evolution of the magnetic inclination angle $\chi$ and the spin frequency $\nu=\Omega/2\pi$ can be determined, and the results are illustrated in Fig. \ref{fig: T_bulk viscosity_Z} (c) and (d) respectively. In Fig. (c), the evolution of the magnetic inclination angle is shown to be particularly sensitive to the Z-factor. Without the Z-factor (depicted by the dotted line), the magnetic inclination angle increases rapidly to 90 degrees in approximately $10^{-5}$ years. However, when accounting for the Z-factor (solid line), $\chi$ firstly increases relatively slowly and then drops to a stable value, because the Z-factor inhibits the bulk viscosity during this stage, resulting in a relatively lower rate of increase in the magnetic inclination angle. As shown in Fig. \ref{fig: T_bulk viscosity_Z} (d), the spin frequency with the Z-factor is higher than that without the Z-factor. That is, the Z-factor retards the spin-down of the magnetar, stemming from the Z-factor reduced $\chi$ and thus the small magnitude of $\Dot{\Omega}$ according to equation (\ref{the evolution of angular frequency}).

The decrease in spin frequency stems from the spin energy loss through GWR and MDR. The luminosities of GWR and MDR are depicted in Fig. \ref{fig: T_bulk viscosity_Z} (e). Before $7.2 \times 10^{-4}$ years, regardless of considering the Z-factor, the magnetar loses its spin energy primarily through GWR, i.e., $L_{{\rm GWR}} > L_{{\rm MDR}}$. Subsequently, the energy release channel is predominantly governed by the MDR. The evolution of GWR and MDR luminosities reveals that both two increase with time in the initial $6.4 \times 10^{-6}$ years followed by a subsequent decrease, but this time is delayed slightly after considering the Z-factor. According to equation (\ref{the luminosity of GW and MDR}), either the MDR luminosity or the GWR luminosity relies on the magnetic inclination angle and the spin frequency of the magnetar. While the spin frequency remains essentially unchanged, the magnetic inclination angle experiences a rapid increase, as shown in Fig. \ref{fig: T_bulk viscosity_Z} (c), leading to a swift increase in the radiative luminosity of the magnetar. However, after this, the spin frequency starts to decrease rapidly, which results in a swift decline in the radiative luminosity because of $L_{\rm MDR} \propto \nu^4$ and $L_{\rm GWR} \propto \nu^6$. Due to the Z-factor effect in retarding the spin-down of the magnetar, the arising time of the luminosity peak is delayed when the Z-factor is included. And it is found that the Z-factor reduces the radiated luminosity of the GWR and MDR by a factor of 4 and 2, respectively. 

The MDR of a millisecond magnetar can be used to heat the ejecta of SLSNe, consequently influencing the evolution of the thermal radiant luminosity of SLSNe. Before the age of $2.1 \times 10^{-5}$ years, the energy injected into SLSNe is diminished due to the suppression of MDR by the Z-factor effect. After this age, however, the Z-factor facilitates the MDR, leading to the brighter thermal radiation of SLSNe compared to the scenarios without the Z-factor, as illustrated in Fig. \ref{fig: T_bulk viscosity_Z} (f). Upon reaching the peak luminosity of MDR, the energy injected into SLSNe declines, resulting in a gradual slowdown in the increase of the thermal luminosity of SLSNe, while the thermal radiation continuously consumes the internal energy of the ejecta. When the gain and loss balance, the thermal radiation luminosity of SLSNe reaches its peak. At this peak, the Z-factor effect enhances the radiation luminosity by approximately three times the original amount, that is, the nucleon-nucleon short-range correlation leads to a higher peak magnitude of SLSNe. In a word, with the initial conditions used in this study, the Z-factor suppresses the thermal emission luminosity of SLSNe in a very short period of time after its birth (less than 1 hour), which is almost impossible to observe, so the Z-factor predominantly elevates the thermal emission luminosity of SLSNe on the whole.

\begin{figure*}
    \centering
    \includegraphics[width=1\linewidth]{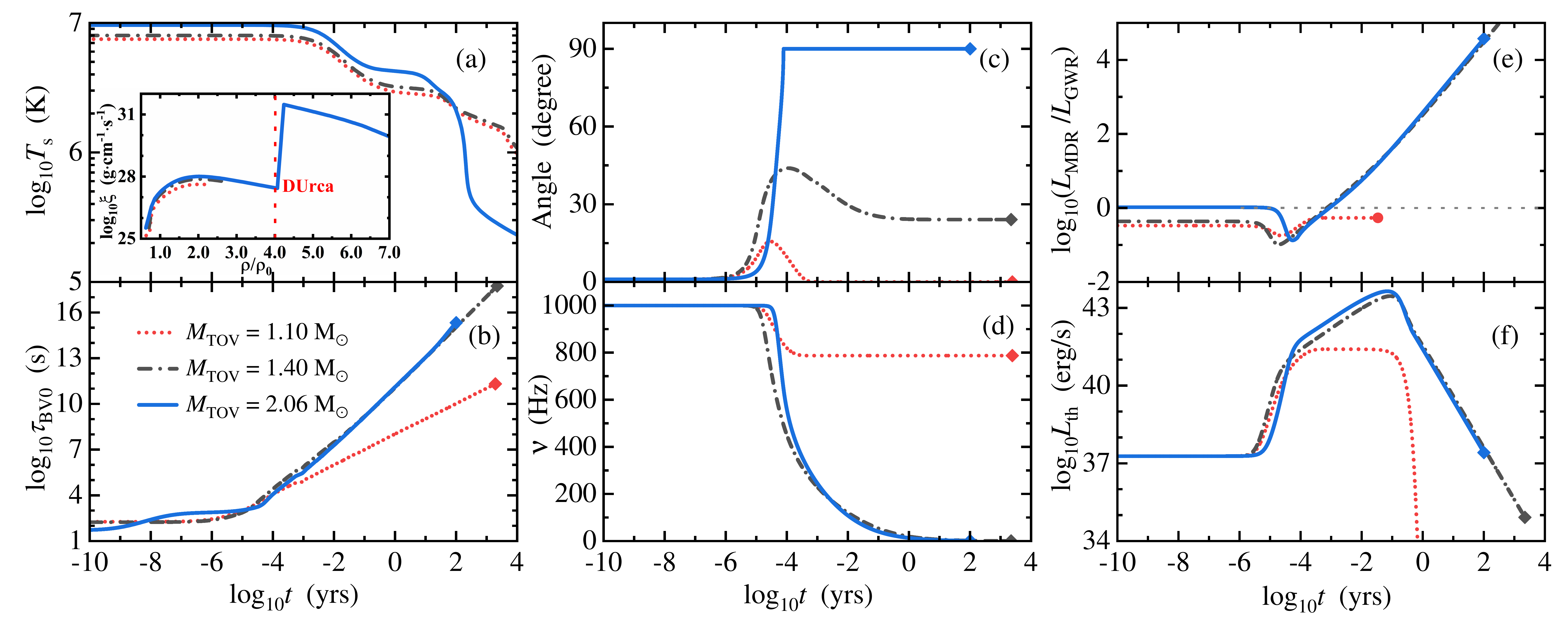}
    \caption{(Color online) The evolution of the stellar surface temperature $T_{\rm s}$, the reduced damping timescale $\tau_{\rm BV0}$, magnetic inclination angle $\chi$, spin frequency $\nu$, GWR and MDR luminosity, and the radiated luminosity of a magnetar with an initial period $P_0=1 $ms under different masses. The symbol for a solid circle stands for $L_{\rm MDR} < 10^{-128} \, \rm erg/s$.}
    \label{fig: Angle_frequency_M}
\end{figure*}

\subsection{Effects of mass and initial spin period}
\label{sec: Effects of mass and initial spin frequency}

Different magnetars have different masses and initial spin periods, and these initial states can also play a significant role in influencing the thermal emission luminosity of SLSNe. In this work, we delve into the effects of magnetar mass and initial spin period on magnetar evolution and the radiative luminosity of SLSNe. We continue to apply the equation of state derived from relativistic mean field theory with FSUGarnet interaction, and the Z-factor effect has been taken into account in all calculations.

When the mass of a magnetar exceeds the critical threshold, the proton fraction in the core is sufficiently high to trigger the DUrca process. The maximum non-rotating mass, $M_{{\rm TOV}} = 2.06 \ {\rm M_\odot}$, is  derived from the FSUGarnet interaction. In the case of such a massive neutron star, the DUrca process occurs in the core when the density is larger than approximately 4 times the saturation density, resulting in a substantial increase in the bulk viscosity of the neutron star matter in this density region by around $3\sim4$ orders of magnitude. The acceleration in the cooling process is a direct consequence of the DUrca process. The DUrca process affects the bulk viscosity in two ways. On the one hand, it directly leads to larger bulk viscosity according to equation (\ref{ bulk viscosity of DU}) since the bulk viscosity caused by the DUrca process is several orders of magnitude larger than that caused by the MUrca processes. On the other hand, it significantly accelerates the neutron star cooling, which results in a lower temperature and smaller bulk viscosity inside the star at a given age on the whole. As shown in Fig. \ref{fig: Angle_frequency_M} (b), the combined effects causes the massive neutron star with $2.06 \, \rm M_{\odot}$ to exhibit a shorter damping time scale at the beginning. Since the DUrca is more efficient than the MUrca, the temperature of the DUrca region of the neutron star interior drops quickly. As the DUrca region is distinctly colder compared with the MUrca region and acts as a heat sink, the cooling rate becomes relatively slower (Sales, Thiago et al. \citeyear{sales2020revisiting}), resulting in slowly increasing the reduced damping timescale. Finally, since $\tau_{\rm BV0} \propto \nu^{-2}$, the dramatically reduced spin leads to a rapid increase in the reduced timescale. For low-mass ($M_{{\rm TOV}} = 1.1 \ {\rm M_\odot}$) neutron star, given that the spin is extremely rapid, the reduced damping timescale is relatively shorter after $10^{-3}$ years.

Compared with the case of a canonical magnetar with $M_{{\rm TOV}} = 1.4 \ {\rm M_\odot}$, the magnetic and rotational axes of low-mass (high-mass) magnetars with $M_{{\rm TOV}} = 1.1 \ {\rm M_\odot}$ ($M_{{\rm TOV}} = 2.06 \ {\rm M_\odot}$) are more readily aligned (orthogonal), as shown in Fig. \ref{fig: Angle_frequency_M} (c). For a neutron star with $1.1 \ {\rm M_\odot}$, as the magnetic inclination angle decreases, both GWR and MDR are suppressed, resulting in a decrease in the spin energy loss rate. When the $\chi $ value reaches zero, the spin energy losses are completely halted, causing the spin frequency evolution to cease, as exhibits in Fig. \ref{fig: Angle_frequency_M} (d). At the beginning, the ratio of MDR luminosity to GWR luminosity increases as the neutron star mass increases, as shown in Fig. \ref{fig: Angle_frequency_M} (e). Since this ratio is proportional to $[\Omega^2(15\sin^2{\chi} + 1) ]^{-1}$, the increase in magnetic inclination angle dominates until this ratio reaches a minimum, and thereafter the spin-down dominates this ratio. In the case of a canonical magnetar, the GWR mainly dominates before $7 \times 10^{-4}$ years, while MDR takes over thereafter. In the case of the $2.06 \ {\rm M_\odot}$ magnetar, the GWR and MDR are evenly matched initially, then GWR gains the upper hand, and finally the MDR dominates gradually. Conversely, for the neutron star with a mass of $1.1 \ {\rm M_\odot}$, GWR dominates in the lost spin energy, and the ratio of MDR luminosity to GWR luminosity is almost a constant after $4 \times 10^{-4}$ years due to the magnetic inclination angle close to 0 degrees. Therefore, the initial spin energy of magnetars is released through GWR shortly after birth (less than several days), which drastically reduces the MDR energy injected into SLSNe, resulting in a substantial reduction in the corresponding radiant luminosity of SLSNe. The peak radiant luminosity of SLSNe powered by low-mass magnetars is approximately 2 orders of magnitude lower than that powered by a $1.4 \ {\rm M_\odot}$ canonical magnetar, as shown in Fig. \ref{fig: Angle_frequency_M} (f). In the magnetar-powered model, once the MDR is fully suppressed, the radiated luminosity of SLSNe experiences a rapid decline after reaching its peak due to adiabatic expansion. In other words, if the magnetic inclination angle of a magnetar in SLSNe quickly reaches 0 degrees, its radiation luminosity also decreases quickly until it becomes unobservable. In contrast, for SLSNe receiving a continuous supply of MDR energy, a gradual decrease in thermal radiation luminosity is evident. Additionally, the peak radiant luminosity of SLSNe elevates as the mass of the magnetar increases.

\begin{figure*}
    \centering
    \includegraphics[width=0.95\linewidth]{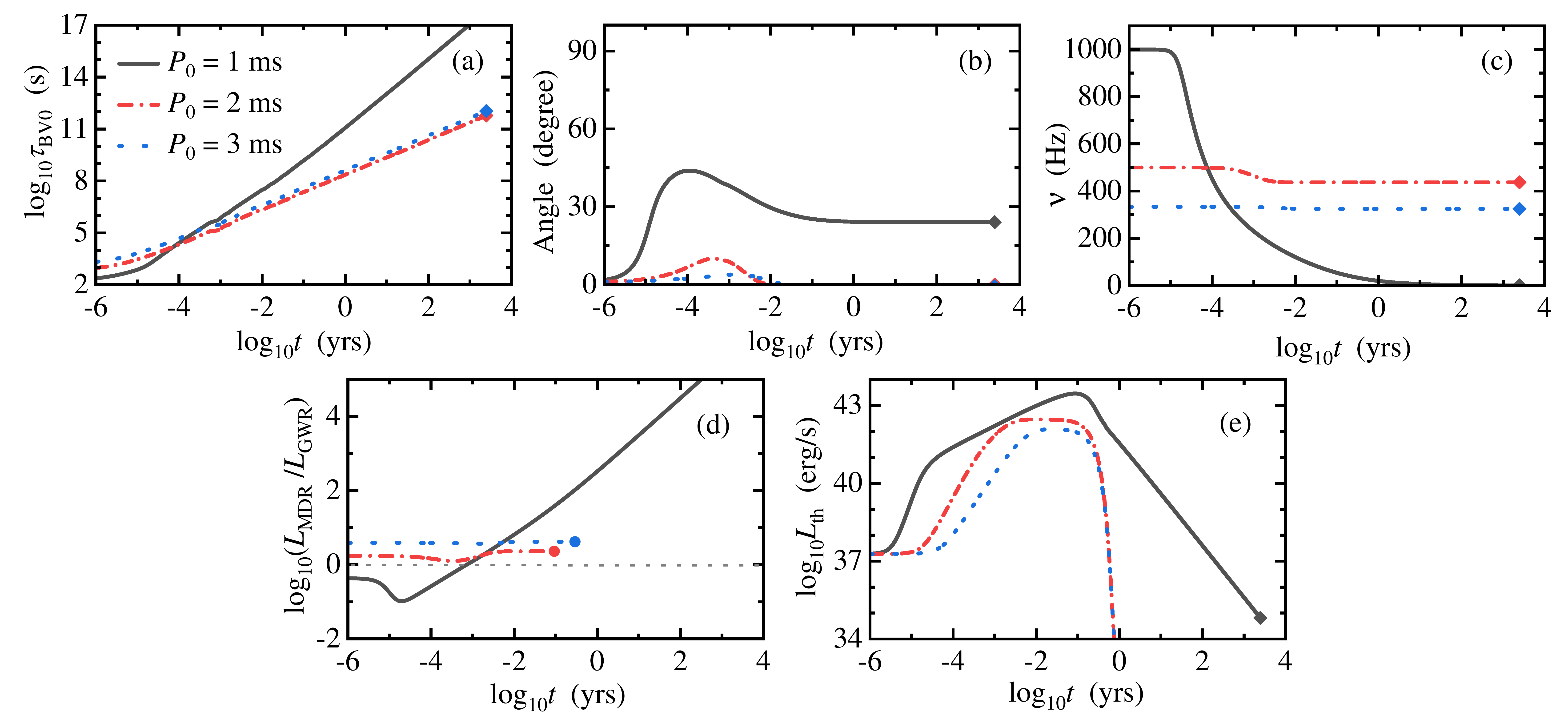}
    \caption{(Color online) The evolution of reduced damping timescale $\tau_{\rm BV0}$, the magnetic inclination angle $\chi$, spin frequency $\nu$, GWR and MDR luminosity and the radiated luminosity of a canonical magnetar under different initial period $P_0 $.}
    \label{fig: Pall}
\end{figure*}

Subsequently, we discuss the impact of the initial spin period of the magnetar on the radiated luminosity. According to equations (\ref{ bulk viscosity of DU} - \ref{the total bulk viscosity}), a slower rotation of the neutron star leads to enhanced interior viscosity and hence a shorter viscosity-damping timescale. In our calculation, the bulk viscosity damping rate is proportional to $\nu^4$ (Dall'Osso et al. \citeyear{dall2009early}), with slower rotation leading to a longer reduced damping timescale, as displayed in Fig. \ref{fig: Pall} (a) and (c). A slow initial rotation free-body precession magnetar is more likely to quench the magnetic inclination angle. As shown in Fig. \ref{fig: Pall} (b), the larger the initial period, the smaller the peak value of the magnetic inclination angle is. A reduced magnetic inclination angle suppresses both MDR and GWR, consequently inhibiting the decrease in spin frequency. For instance, in the case of neutron stars with initial spin periods of $P_0 = 2\ \rm ms$ and $P_0 = 3 \ \rm ms$, as shown in Fig. \ref{fig: Pall} (c), the spin frequency decreases slightly in the later stages of their evolution, and the magnetar loses its spin energy mainly through MDR throughout the evolutionary process, as shown in Fig. \ref{fig: Pall} (d). Particularly for neutron stars with very slow initial spins (i.e. $P_0 = 3 \ \rm ms$), the spin frequency remains nearly unchanged. This results in the suppression of both MDR and GWR, and therefore less energy is injected into SLSNe, leading to a significant reduction in radiated luminosity, as shown in Fig. \ref{fig: Pall} (e). We may conclude that the SLSNe is brighter when the initial rotation of the magnetar is faster in the considered magnetic field configuration of the present study, highlighting the critical role of the initial spin period in the luminosity of SLSNe.

\subsection{Effect of symmetry energy on the evolution of newborn magnetars and SLSNe}
\label{sec: Effect of symmetry energy on the evolution of newborn magnetars and SLSNe}

\begin{figure}
    \centering
    \includegraphics[width=1\linewidth]{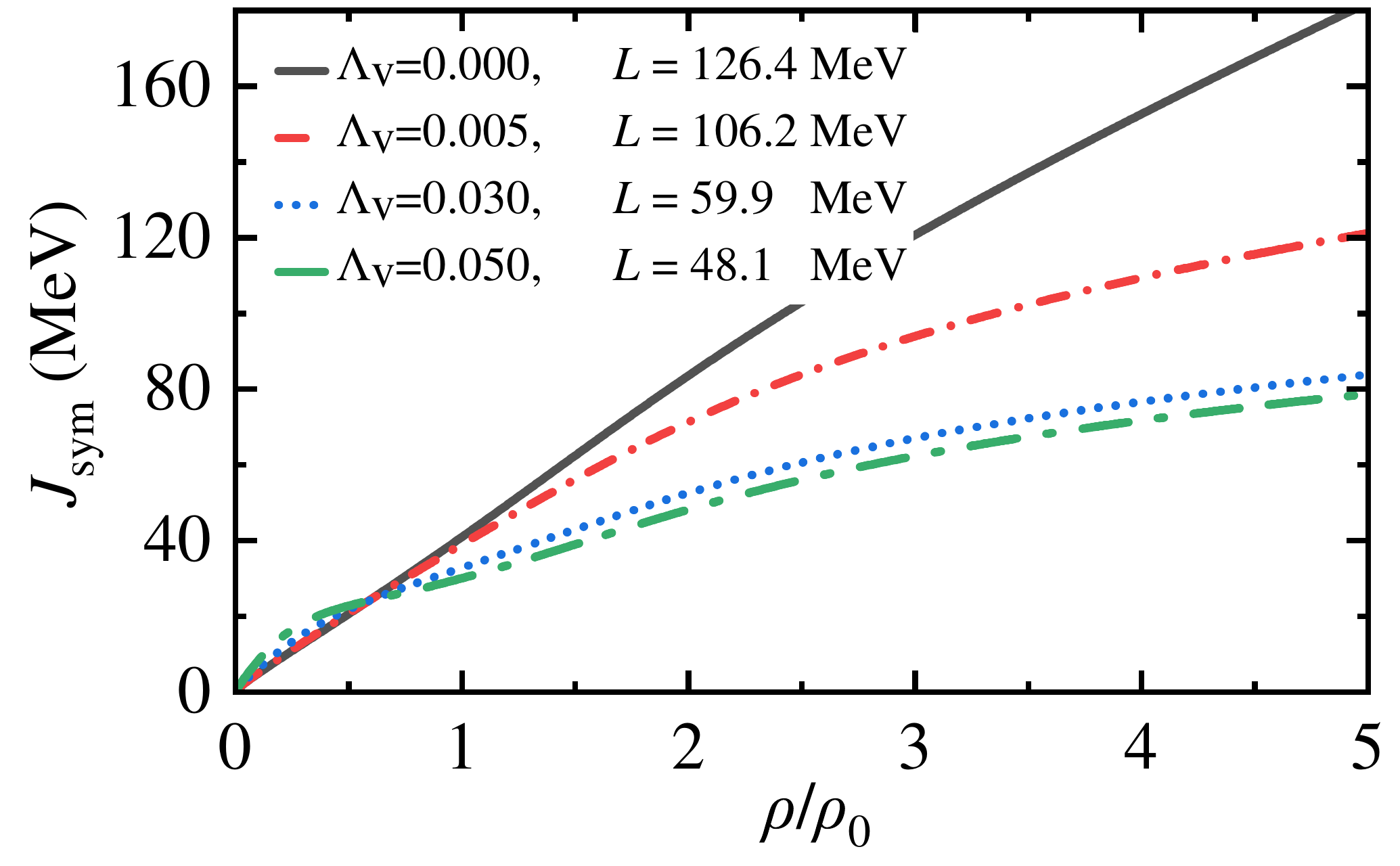}
    \caption{(Color online) Symmetry energy in units of the saturation density as a function of density for different $\Lambda_{\text{V}}$ parameter in the family of the FSUGarnet interaction. }
    \label{fig: FSUGarnet_Esym}
\end{figure}
\begin{table}
	\centering
	\caption{Parameter sets of the FSUGarnet family. The symmetry energy $J$ and the corresponding slope parameter $L$ in units of MeV at the saturation density $\rho_0$ are presented.}
	\label{tab: EoS}
        \scalebox{1.3}{
        \begin{tabular}{lccr} 
		\hline
		$\Lambda_V$ & $g_\rho^2$ & $J$ & $L$\\
		\hline
		0.000 & 88.7450 & 41.0 & 126.4\\
            0.005 & 94.6357 & 39.1 & 106.2\\
		0.030 & 141.6465 & 32.9 & 59.9\\
        0.050 & 235.0607 & 30.1 & 48.1\\
		\hline
	\end{tabular}
        }
\end{table}

We examine the role of density-dependent symmetry energy of dense nuclear matter in the evolution of magnetar based on the relativistic mean-field model with FSUGarnet interaction. 
The symmetry energy which characterizes the isospin dependence of the equation of state of asymmetric nuclear matter, plays a crucial role in a number of intriguing issues of nuclear physics and astrophysics. 
The equation of state of nuclear matter can be written as $\varepsilon(\rho,\beta) = \varepsilon(\rho,0) + J_{\rm sym}(\rho )\beta^2 + o(\beta^4)$ with the density $\rho = \rho_n + \rho_p$ and isospin asymmetry $\beta = (\rho_n - \rho_p)/\rho$. $\varepsilon(\rho,0)$ is the equation of state of the symmetric nuclear matter with $\rho_n = \rho_p$ which is well constrained by heavy-ion experiments. The FSUGarnet interaction is used here because it provides a good description of the equation of state of symmetric nuclear matter (Fattoyev et al. \citeyear{fattoyev2020gw190814}). The symmetry energy $J_{\rm sym}(\rho) $ is highly uncertain. The uncertainly in the equation of state is mainly attributed to $J_{\rm sym}(\rho) $.
In the framework of relativistic-mean field theory, it is given by
\begin{align}
    J_{\rm sym}(\rho) = \frac{k_{\rm F}^2}{6\sqrt{k_{\rm F}^2 + (m-g_\sigma \sigma)^2}} + \frac{g_{\rm \rho}^2}{12\pi^2}\frac{k_{\rm F}^3}{(m_\rho^2 + 2 \Lambda_Vg_\rho^2g_\omega^2 \omega^2)^2} ,
\end{align}
where $k_F$ is the Fermi momentum of symmetric matter. At the saturation density $\rho_0$, the slope parameter of symmetry energy $J = J_{\rm sym}(\rho_0)$ is given by $L = 3\rho_0 (dJ_{\rm sym}/d\rho) \mid_{\rho  = \rho_0} $, which characterize the density dependence of the symmetry energy around saturation density $\rho_0$. Since the $L$ value is very sensitive to $\Lambda_V$, we change the stiffness of the symmetry energy by modifying the value of $\Lambda_V$. Since the uncertainly in the equation of state mainly stems from the symmetry energy term as mentioned above, we introduce a “family” of FSUGarnet by adjusting the isovector parameters $\Lambda_{\text{V}}$ and $g_{\rho}$ in such a way that the value of the symmetry energy remains fixed at a baryon density of $\rho =0.09$ fm$^{-3}$, with the same procedure as in (Piekarewicz \citeyear{PhysRevC.83.034319}) for a “family” of FSUGold models, keeping $\varepsilon(\rho,0)$ unchanged. There is inconsistency in using the relativistic mean-field model for the equation of state and the Brueckner approach to parametrize the density dependence of the Z-factors. When calculating the equation of state by using the microscopic Brueckner method, the symmetry energy cannot be adjusted freely as in the phenomenological relativistic mean-field model, keeping the properties of the symmetric nuclear matter unchanged. On the other hand, the relativistic mean-field model does not take into account nucleon-nucleon correlation effects, and thus it cannot be employed to compute the Z-factor. The equation of state is used to build the stellar structure, and the Z-factor is used to correct the transport properties of dense matter. This inconsistency is unavoidable at present but does not affect our qualitative conclusions. 

In this work, we offer the symmetry energy from stiff ($\Lambda_{\text{V}}=0.000$) to soft ($\Lambda_{\text{V}}=0.050$), and we display in Fig. \ref{fig: FSUGarnet_Esym} the symmetry energy predicted by all of the adjusted interactions as a function of density. The symmetry energy is divergent distinctly at high densities. We listed in Table \ref{tab: EoS} the explicit parameters $\Lambda_{\text{V}}$ and $g_{\rho}$ for the FSUGarnet family together with the symmetry energies and their slope parameters $L$. The smaller the coupling constant $\Lambda_V$, the stiffer the symmetry energy is. 
\begin{figure*}
    \centering
    \includegraphics[width=1\linewidth]{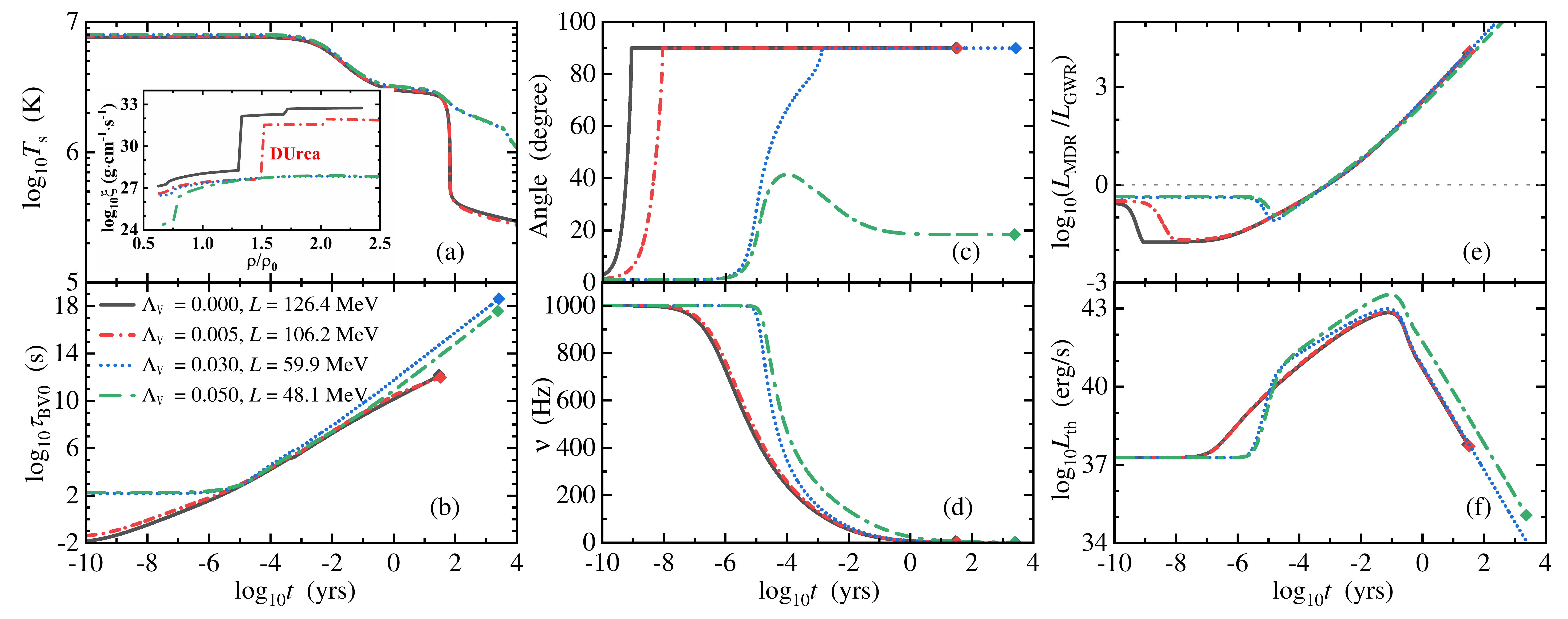}
    \caption{(Color online) The evolution of stellar surface temperature $T_{\rm s}$, reduced damping timescale $\tau_{\rm BV0}$, magnetic inclination angle $\chi$, spin frequency $\nu$, GWR and MDR luminosity, and the radiated luminosity of a canonical magnetar with initial period $P_0=1 $ms under different behavior of the density-dependence of symmetry energy.}
    \label{fig: Angle_frequency_EOS}
\end{figure*}
 
The density-dependence of the symmetry energy determines the fraction of each component in the neutron star core, and it is directly linked to the bulk viscosity coefficient as outlined in equations (\ref{ bulk viscosity of DU} - \ref{bulk viscosity of MU pu}), but its effect is difficult to disentangle. The numerically calculated results about the roles of symmetry energy in the cooling process, reduced viscosity timescale, spin evolution, and magnetic inclination angle evolution of a canonical magnetar with the mass of $1.4 \ {\rm M_{\odot}}$ are presented in Fig. \ref{fig: Angle_frequency_EOS}. In the case of $\Lambda_V = 0.000$ and $\Lambda_V = 0.005$, the proton fraction is high enough to trigger the DUrca process, leading to a notable enhancement of the bulk viscosity coefficient. As the bulk viscosity of dense matter inside the star gets stronger, the damping effect on the free-body procession becomes stronger, and consequently the magnetic inclination angle $\chi $ is more likely to evolve to 90 degrees. As shown in Fig. \ref{fig: Angle_frequency_EOS} (c), in the case of $\Lambda_V = 0.000$, $\Lambda_V = 0.005$, and $\Lambda_V = 0.030$, the orthogonal of the spin and magnetic axes is achieved ($\chi =90$ degrees) in a very short time, significantly enhancing GWR and MDR. Thereafter, the spin energy experiences a substantial decrease, leading to a rapid spin-down of the magnetar. As shown in Fig. \ref{fig: Angle_frequency_EOS} (d), the stiffer the symmetry energy, the earlier the spin frequency begins to drop significantly. Additionally, in Fig. \ref{fig: Angle_frequency_EOS} (e), it is demonstrated that compared to a soft symmetry energy a stiffer symmetry energy results in a stronger GWR relative to MDR energy before $\sim  10^{-5}$ years. Regardless of whether the symmetry energy is soft or stiff, most of the spin energy is dissipated through GWR, and a stiff symmetry energy is more favorable for GWR. After $3 \times 10^{-3}$ years, the loss of the remaining little spin energy is primarily driven by MDR.

For $\Lambda_V = 0.000$ and $\Lambda_V = 0.005$, since the spin frequency exhibits a significant decrease in the early phase, the thermal emission luminosity of SLSNe is much higher than the other cases between $10^{-7} \sim 10^{-5} $ years, as shown in Fig. \ref{fig: Angle_frequency_EOS} (f). However, as discussed above, the majority of the spin energy is dissipated via GWR at this stage. Therefore, after $10^{-5}$ years, less energy is injected into the SLSNe compared to the cases of $\Lambda_V = 0.03$ and $\Lambda_V = 0.05$. As the MDR energy continues to heat the ejecta, the peak value of the thermal radiative luminosity obtained from the stiff symmetry energy is lower than that from the softer symmetry energy. The peak luminosity of the thermal radiation for the $\Lambda_V = 0.050$ case is 5 times larger than that for the $\Lambda_V = 0.000$ case. 
The radiative luminosity of the $\Lambda_V = 0.005$ and $\Lambda_V = 0.000$ is basically a perfect match. Only if the symmetry energy is sufficiently stiff, the effect of symmetry energy change on SLSNe is negligible. 
Therefore, the symmetry energy affects the radiative luminosity of SLSNe visibly. The softer the symmetry energy, the higher the peak radiative luminosity is.

The density dependence behavior of the nuclear symmetry energy at high densities is still poorly known (Li et al. \citeyear{Li2019Towards}; Baldo \& Burgio \citeyear{BALDO2016203}). In terrestrial laboratories, it is constrained by heavy-ion collisions (Lattimer \& Steiner \citeyear{Lattimer2014}). Recently, astronomical observations such as neutron star merger and mass-radius measurements, have been extensively employed to further constrain the symmetry energy at higher densities (Abbott et al. \citeyear{PhysRevLett2017GW170817}; Abbott et al.  \citeyear{PhysRevLett2018GW170817}; Miller et al. \citeyear{Miller2019}; Tanvir et al. \citeyear{Tanvir2013}). The observed magnetic inclination angle distribution of neutron stars may offer a novel strategy to explore the symmetry energy at high densities. The relevant research is currently underway, which help one to gain further insights into the fundamental aspect of nuclear physics.

\section{Summary}
\label{sec: Summary}
Millisecond magnetars are regarded as potential central engines for some intriguing astronomical phenomena such as Hydrogen-poor SLSNe. Taking into account the damping of free-body precession due to the bulk viscosity of neutron star matter, we have studied the first stage of the evolution of magnetic inclination angle and spin of the newly formed and rapidly rotating magnetars (with some typical parameters $B_d = 6 \times 10^{14} \, \rm G$, $\Bar{B}_t = 5.8 \times 10^{16} \, \rm G$, $P_0 =1\, \rm ms$, $T(r=0, t =0) = 10^{10}\, \rm K$ and $M_{\rm TOV} = 1.4 \, \rm M_\odot$) along with the corresponding evolution of the thermal radiation luminosity of SLSNe. 
We delve into the effects of nucleon-nucleon short-range correlations  (i.e., the $Z$-factor effect), the initial conditions of magnetars, and the density-dependent symmetry energy on this evolution. A key physical quantity we calculated is the time- and space-dependent bulk viscosity coefficient of dense matter and the corresponding timescale. This bulk viscosity is able to drive the magnetic axis orthogonal to the spin axis while the GWR leads to alignment of the spin and magnetic axes (i.e., reduces the magnetic inclination angle $\chi$). The equation of state, as an input for the calculation of bulk viscosity coefficient and neutron star structure, is obtained from the relativistic mean-field theory with the FSUGarnet interaction.

The Z-factor effect directly reduces the bulk viscosity and neutrino emissivity. On the other hand, it slows down the cooling of young neutron stars and thus indirectly enhances the bulk viscosity of dense matter inside the stars. 
Besides these two mechanisms, the free-body precession of the neutron star also affects the bulk viscosity given that magnetic inclination angle, angular frequency, and bulk viscosity are coupled together. Consequently, the effect of the Z-factor on the evolution of the bulk viscosity is dependent on the initial state of the magnetar.
However, due to the Z-factor directly decreasing the bulk viscosity at the beginning, for arbitrary initial states, 
the Z-factor is found to suppress the growth in the magnetic inclination angle. Moreover, it retards the spin energy loss. In other words, with the parametric used in this study, the Z-factor suppresses both MDR and GWR luminosity until $\sim 10^{-5}$ years old and thereafter enhances the luminosity. The peak luminosities of MDR and GWR are reduced by factors of 2 and 4, respectively. The spin energy loss of the magnetar is mainly through GWR (about $7.2 \times 10^{-4}$ years ago) and then through MDR. As the Z-factor enhances the MDR after 1 hour, it facilitates more MDR energy to heat the SLSNe ejecta and thus triples the peak value of thermal radiative luminosity of SLSNe.

The mass and initial period of a millisecond magnetar must fall within specific ranges to ensure that its luminosity peak reaches at least the lowest luminosity peak ($10^{43} \ \rm erg/s$) of the SLSNe (Chen et al. \citeyear{chen2023hydrogen}; Tinyanont et al. \citeyear{tinyanont2023supernova}). Generally, the brightness of SLSNe reaches its peak in a few tens of days. For some magnetars, the magnetic and rotational axes take a very short time to reach alignment, just a few days. Consequently, the MDR is almost completely quenched before the luminosity of SLSNe reaches its peak, ultimately leading to the inability of such magnetars to act as a central engine for SLSNe. For instance, under the magnetic field conditions in this study, canonical magnetars with relatively slow rotation of $P_0 \ge 2 \, \rm ms $ and with small masses are perhaps unable to provide enough energies to act as the central engine of SLSNe, although the spin energy is released through MDR for magnetars with slower initial spins. Other interesting results have been found. For instance, for magnetars with slow initial spins and small mass, the magnetic inclination angle quickly evolves to 0 degrees, and the spin remains essentially unchanged in the late stages of evolution.
 
The evolution of magnetars is highly sensitive to the density-dependent symmetry energy. To explore this symmetry energy effect, we employ the relativistic mean-field model with the “family” of FSUGarnet interactions which exhibit different density-dependent behavior. In this “family” of the equation of state, our findings reveal that the stiffness of the symmetry energy directly impacts the bulk viscosity of the star. When the proton fraction inside the star is high enough to reach a threshold, the DUrca process is triggered, which significantly increases the bulk viscosity by several orders of magnitude. Consequently, the magnetic inclination angle rapidly increases to 90 degrees, which greatly facilitates GWR. Furthermore, with this “family” of FSUGarnet interaction, the peak luminosity of SLSNe decreases as the symmetry energy becomes stiffer. Based on the factors analyzed above, it has been also found that if the magnetic inclination angle is not 0 degrees in the first stage, the spin energy is lost dominantly through GWR about 1 day ago, and later predominantly through MDR. 

This study focuses on the first evolution stage (non-superfluid stage) of the magnetic inclination angle evolution. To achieve a more complete understanding of the magnetar evolution and hence the magnetar-powered SLSNe light curves, further investigation into the superfluid stage is essential. The superfluid stage is much more complex than the first evolution stage, and the shear viscosity and the crust-core coupling may need to be addressed. In addition, the GWR induced by $r$-mode instability is perhaps non-negligible for the magnetar evolution and the light curve of SLSNe. These researches are in progress.

\section*{Acknowledgements}
This work is supported by the National Natural Science Foundation of China under Grants No. 12222511, by the Chinese Academy of Sciences Project for Young Scientists in Basic Research YSBR-088, by the Strategic Priority Research Program of Chinese Academy of Sciences, Grant No. XDB34000000, and by the Continuous Basic Scientific Research Project under Grants No. WDJC-2019-13.

\section*{Data Availability}
The data used to support the findings of this study are available from
the corresponding author upon request.



\bibliographystyle{mnras}
\bibliography{reference} 








\bsp	
\label{lastpage}
\end{document}